\documentclass[preprint,preprintnumbers,amsmath,amssymb]{revtex4}
\usepackage{graphicx}
\usepackage{color}
\usepackage{times}
\usepackage{amssymb}
\usepackage{amsmath}
\usepackage{latexsym}
\usepackage{graphicx}
\usepackage{dcolumn}
\usepackage{bm}
\usepackage{amsthm,amsfonts} 

\def\TJ{\tilde J}

\def\R{\mathbb{R}}
\def\N{\mathbb{N}}
\def\C{\mathbb{C}}

\def\Z {\mathbb{Z}}

\def\lg{\langle }
\def\rg{\rangle }
\newcommand\deq{\stackrel{\mathrm{def}}{=}}

\def\TJ{\tilde{J}}

\begin{document}
\title{Quantization with Action-Angle Coherent States}


\author{Jean Pierre Gazeau}
\affiliation{Astroparticules et Cosmologie, Univ Paris Diderot, Sorbonne Paris Cit\'e, 75205 Paris-Fr }
\email{gazeau@apc.univ-paris7.fr}
\author{Rina Kanamoto}
\affiliation{Division of Advanced Sciences, Ochanomizu University, 2-1-1 Ohtsuka, Bunkyo-ku, Tokyo 112-8610-Jp}
\email{kanamoto.rina@ocha.ac.jp}


\begin{abstract}
For  a single degree of freedom confined mechanical system with given energy, we know that the motion is always periodic and action-angle variables are convenient choice as conjugate phase-space variables. We construct action-angle coherent states in view to provide a quantization scheme that yields precisely a given observed energy spectrum $\{E_n\}$ for such a system. This construction is based on a Bayesian approach: each family corresponds to a choice of probability distributions such that the classical energy averaged with respect to this probability distribution is precisely $E_n$ up to a constant shift. The formalism is viewed as a natural extension of the Bohr-Sommerfeld rule and an alternative to the canonical quantization. In particular, it also yields a satisfactory angle operator as a bounded self-adjoint operator. 
\end{abstract}
\maketitle

\section{Introduction}
 \label{intro}

\subsection*{Action-angle variables}

 Let us consider a a single degree of freedom confined mechanical system described with phase-space conjugate variables  $(q,p)$. Suppose it conservative. For a given motion its Hamiltonian function is fixed to a certain value $E$ of the energy $H(q,p) = E$.
Solving for the momentum variable $p$ leads to $p=p(q,E)$.
We suppose that we are in presence of a confinement of the system where we have periodic motions only. 
Then two types of periodic motions are possible:
\begin{itemize}
 \item[(i)] {\it Libration}: phase trajectory is closed, and then $q$ and $p$ are periodic functions of time with a same period. 
\item[(ii)] {\it Rotation} or {\it circulation}: phase trajectory is not closed, then $p$ is periodic function of $q$. 
\end{itemize}
For either type, introduce  the \emph{action variable} \cite{gold}
\begin{equation}
\label{action1}
J = \oint p(q,E)\, dq = J(E)\, , 
\end{equation}
where the loop integral is understood as performed over a complete period of libration (resp. rotation).
This determines by inversion of $J= J(E)$ the function $E = E(J)$. Now we know that the Hamilton-Jacobi equation $\frac{\partial S}{\partial t} + H\left(q, \frac{\partial S}{\partial t}\right)=0$
obeyed by the action $S= \int L \, dt$,  with $L= p\, \dot q -H$, has a solution of the type 
$S= W(q,J) -E\, t$ (note that the action $S$ should not be confused with the action variable $J$).
The time-independent $W=W(J,q) = \int p\,dq $ is the \emph{Hamilton characteristic function} which generates the contact transformation $(q,p)\mapsto (J,\gamma)$ at constant Hamiltonian, where $\gamma = \frac{\partial W}{\partial J}$ is the angle variable, conjugate to  $J$. 
It follows from the definition of $J$ that the period $\tau$ (resp. frequency $\nu$) of motion at fixed energy is $
\tau= \partial J/\partial E = \tau (E)$ (resp. $\nu(E)= 1/\tau(E)= \partial E/\partial J$), and so the  time evolution of the angle variable is linear, with period $\tau$:
\begin{equation}
\label{angle}
\gamma = \frac{t}{\tau(E)} + \gamma_0 = \nu(E)\, t + \gamma_0\,. 
\end{equation}
 Note that this equation allows to consider the time as proper to the system at given energy. 
 
\subsection*{Action-angle coherent states for measured energies}

 Let us suppose that a series of energy measurements  on a  mechanical system with one-degree of freedom yields an energy spectrum $E_0, \, E_1, \, \dotsc E_n, \, \dotsc$. In this paper families of corresponding action-angle coherent states  are constructed  in view to provide a quantization scheme consistent with that discrete sequence of experimental energies.  The construction is based on a Bayesian approach: each family corresponds to a choice of probability distributions,  $n\mapsto p_n(J)/\mathcal{N}(J)$ (prior discrete), $J\mapsto p_n(J)$ (posterior continuous) such that the mean value of the classical energy with respect to the probability $p_n(J)$ is precisely $E_n$, up to the addition of a constant independent of $n$. 
 The formalism \cite{ghk11} can be viewed as a natural extension of the empirical Bohr-Sommerfeld rule, $J(E)= nh$,
 where $h$ is the Planck constant. We  know that this quantization is exact for the motion on the circle (quantization of the angular momentum), and valid in the semiclassical regime. In the deep quantum regime our approach can be viewed as a viable alternative to the canonical quantization, particularly when the latter is impracticable. For instance,  it yields a satisfactory angle operator.
 
 In Section \ref{csfamily} we give an overview of a general construction of coherent states labelled by elements of a measure set,  their use as a quantizer frame, and their suitability in regard with their phase space content.  
We specify this formalism  in Section \ref{bayes} to the action-angle phase space of a confined system by following a Bayesian construction  mentioned above. 
    In Section \ref{expprob} we revisit the question of  suitability of these action-angle coherent states. Sections \ref{partprob} and \ref{genprob} are devoted to the question of choice of probability distributions appropriate for our quantization goals. Section \ref{timevfr} is an illustration of our approach by examining a few characteristics, like time evolution,  of two types of coherent states for the free rotator.  We end this note in Section \ref{concl} with some comments about  questions raised by our approach and possible generalizations. 


\section{Coherent states family as Hilbertian frame for phase space}
 \label{csfamily}
  
 \subsection{The general construction \cite{jpgbook09}}
 
 Let $X$ be the phase space of a mechanical system, equipped with its symplectic measure $\mu$. Actually $(X,\mu)$ can be any measure space.
 Let $\mathcal{I}$ be some countable set and  $\mathcal{O}$ be  an orthonormal system $\mathcal{O}= \{\phi_n \, , n \in \mathcal{I}\}$ made of elements $\phi_n(x)$  in the Hilbert space $L_{\C}^2(X , \mu)$, with  the positiveness and finiteness constraints
\begin{equation}
\label{finconst }
0< \mathcal{N}(x) \deq  \sum_n \vert \phi_n(x)\vert^2  < \infty \quad \mathrm{a.e.}\, x \in X\, . 
\end{equation}
Let $\mathcal{H}$ be some separable Hilbert space with orthonormal basis $\{|e_n\rg\, , n \in \mathcal{I}\}$ in one-to-one correspondence with elements of $\mathcal{O}$: $ |e_n\rg  \mapsto \phi_n$. 
Then the following family of states in the companion $\mathcal{H}$, labelled by elements of $X$,
\begin{equation}
\label{csgen}
|x\rg = \frac{1}{\sqrt{\mathcal{N}(x)}}\sum_{n} \overline{\phi_n(x)}|e_n\rg\, , 
\end{equation}
where $\overline{\phi_n(x)}$ refers to the complex conjugate, and 
obey normalization,  $\lg x | x\rg = 1$, and resolution of the unity $\int_{X}\mu(dx)\,\mathcal{N}(x)\, |x\rg\lg x| = 1_{\mathcal{H}}$. Due to these two properties, vectors $|x\rg$ are named coherent states (CS) (in a wide sense). 
 The resolution of the unity 
\textbf{\large is} precisely the departure point for the corresponding CS quantization 
of functions (or distributions  when the latter are properly defined) on $X$ which transforms them into linear operators in  $\mathcal{H}$:
\begin{equation}
\label{CSquantgen}
f(x) \mapsto A_f = \int_{X}\mu(dx)\,\mathcal{N}(x)\, f(x)\, |x\rg\lg x| \, . 
\end{equation}
$f(x)$ may be considered as  CS quantizable if $A_f$ is densely defined in  $\mathcal{H}$ or if the so-called {\it lower symbol} of $A_f$, $\check{f}(x)\deq \lg x |A_f|x\rg$ is a smooth function on $X$ viewed as a topological phase space. 
Hence, the family $\{|x\rg\}$, $x\in X$ of coherent states offers a certain point of view or frame (in the right Hilbertian geometrical sense) to analyze, in  a non-commutative way, the classical set of points $X$. Changing the family corresponds to a change of frame and possibly to an equivalent or  non equivalent quantization in regard with specific physical quantities, like energy, action, angle, position..., and their mutual Poisson brackets.  

 \subsection{ What we understand by suitable coherent states in Quantum Mechanics}
 Besides the fundamental quantizer role played by the CS $|x\rg$, further criteria are usually requested to 
 qualify the latter as suitable from the  classical-quantum relation point of view. We list here a few of them. For this purpose, we suppose that the CS $|x\rg$ depend on a parameter, say $h$,  the Planck constant, such that the classical limit corresponds to $h \to 0$.

\begin{itemize}
\item[(i)] \textbf{Relative error criterion.}
For a semi-bounded quantizable function $f(x)$ we define the relative error:
\begin{equation}
\label{errorgen}
\mathrm{rerr}_f(x) = \left\vert \frac{\check{f}(x) - f(x)}{f(x) + C}\right\vert\,  , \quad \check{f}(x)\deq \lg x |A_f|x\rg\, ,
\end{equation}
where $C$ is a constant which is  
chosen such that $\vert f(x) + C\vert \neq 0$ for all $x$. Hence: 
\noindent\emph{a CS family is said suitable as a family of quasi-classical states with respect to the function $f$ if $\sup_x\mathrm{rerr}_f(x) \to 0$ as $h \to 0$. }

\item[(ii)] \textbf{Time evolution criterion in semi-classical regime.}
Let $H(x) $ be the Hamiltonian of a mechanical system with phase space $X$ and let $A_H$ be  its CS-quantized counterpart. We define  the time evolution of the  probability density of a CS $|x_0\rg$ in  phase space  with respect to the measure $\mu(dx)$ as the function
\begin{equation}
\label{evphasdist}
\rho_{x_0} (t,x) \deq  \,\mathcal{N}(x)\vert \lg x | e^{-i  A_H t/\hbar}|x_0\rg\vert^2\,. 
\end{equation}
Its probabilistic nature is directly derived from the resolution of the unity. Hence: 
\noindent\emph{a CS family is said suitable as a family of quasi-classical states with respect to the time evolution ruled by the quantum Hamiltonian $A_H$  if the support of the function $\rho_{x_0} (t,x)$ tends to locate on the classical phase space trajectory $x(t)$ with initial condition $x(0) = x_0$ as $h \to 0$.}

\item[(iii)] \textbf{Time evolution stability.} 
\noindent\emph{ A CS family is said temporally stable if under the action of the evolution operator $e^{- i A_H t/\hbar}$, a coherent state is transformed into another coherent state in the same family, up to possibly a phase factor:}
\begin{equation}
\label{tevstab}
e^{-i   A_H t/\hbar}|x\rg = e^{i\beta(t)} |x(t)\rg\, . 
\end{equation}
For more details, see \cite{gk99} and references therein. 
\end{itemize}

 \section{A Bayesian  probabilistic \cite{agh08} construction of action-angle coherent states and related quantizations}
 \label{bayes}
  \subsection*{Conditional posterior probability distribution}
Suppose that  measurement on a confined one-dimensional system yields the sequence of  values for the energy observable (up to a constant shift):
\begin{equation} \label{seqen}
E_0 < E_1< \cdots < E_n < \cdots\,.
\end{equation}
Supposing a (prior) \emph{uniform} distribution on the range of the action variable $J$,  we  define a corresponding sequence of probability distributions $J\mapsto p_{ n}(J)$, i.e. $\int_{\R \,  \mathrm{or}\,\R^+}d\TJ\, p_n(J) = 1$,  with $\tilde{J}\deq J/h$, 
obeying the two  conditions:
\begin{equation} \label{twocond}
  0< \,\mathcal{N} (J) \deq \sum_{n\in \Z\,  \mathrm{or}\, \N}  p_n(J) < \infty\, , \quad E_n + \mathrm{cst}= \int_{\R \,  \mathrm{or}\,\R^+ }d\TJ\, E(J)\, p_n(J) \, ,
 \end{equation}
where $\R$ and $\Z$ (resp. $\R^+$ and $\N$) stand for the rotation (resp. libration)  type of motion. 
The finiteness condition allows to consider the map $n \mapsto p_n(J)/\mathcal{N} (J)$ as a probabilistic model referring to the discrete data, which might be viewed in the present context as a   \emph{prior distribution} also.

 \subsection*{Action-angle coherent states}
  
  Let $\mathcal{H}$ be a complex separable Hilbert space with orthonormal basis 
  $\{|e_{n}\rg\,,\,  n \in \Z \, \mbox{or} \, \N\}$.
  $\mathcal{H}$ is the  space of quantum states. 
  Let $\tau >0$ be a rescaled period of the angle variable and  $X = \{(J,\gamma)\, , \, J \in \R\, \, \mbox{or} \, \R^+, \, 0\leq \gamma < \tau\}$ be the action-angle phase space 
for a rotation (resp. libration) motion 
with given energies  
the discrete sequence $E_0 < E_1< \cdots < E_n < \cdots\,.$
Let  $\left( p_n(J)\right)_{n \in \Z \, \mbox{or} \, \N}$ be the sequence of probability distributions   associated with these energies.  
We  suppose $p_{-n} (J) = p_{n} (-J)$ in the rotation case. 
One then constructs the family of states in $\mathcal{H}$  for the rotation or libration  motion 
as the following continuous map from $X$ into $\mathcal{H}$:
\begin{equation} \label{}
X \ni (J,\gamma) \mapsto |J,\gamma\rg = \frac{1}{\sqrt{\mathcal{N}(J)}}\sum_{n} \sqrt{p_n(J)}\, e^{-i\alpha_n\, \gamma}\, |e_n\rg \in \mathcal{H}\,, 
\end{equation}
where the choice of the real sequence $ n \mapsto\alpha_n$ is left to us in order to comply with some if not all criteria previously listed.

\subsection*{Fundamental properties of action-angle coherent states}
 
 In both cases the coherent states $|J,\gamma\rg$ 
\begin{itemize}
  \item[(i)] are unit vector : $\lg J,\gamma| J,\gamma\rg = 1$
  \item[(ii)] resolve the unity operator in $\mathcal{H}$ with respect a measure ``in the Bohr sense'' $\mu_B(dJ\,d\gamma)$ \cite{gk99} on the phase space $X$ :
  \begin{equation} \label{bohr}
\int_{X} \mu_B(dJ\,d\gamma)\, \mathcal{N}(J)\,| J,\gamma\rg\lg J,\gamma|\deq \int_{-\infty}^{+\infty} d\TJ\, \mathcal{N}(J)\,\lim_{T \to\infty}\frac{1}{T}\int_{-\frac{T}{2}}^{\frac{T}{2}} d\gamma | J,\gamma\rg\lg J,\gamma| =1_{\mathcal{H}}\, ,
\end{equation}
Here, we impose   $T= 2M \tau$ with $M\in \N$ and letting $M\to \infty$. Hence, if the sequence $(\alpha_n)$ assumes all its values in $2\pi\Z/\tau$, then the angular integral reduces to $\frac{1}{\tau} \int_{0}^{\tau} d\gamma$ and the measure $\mu_B$ becomes the ordinary one on the cylinder.
   \item[(iii)] allow a ``coherent state quantization'' of classical observables $f(J,\gamma)$, 
  \begin{equation} \label{quantf(x)}
f(J,\gamma) \mapsto \int_{X} \mu_B(dJ\,d\gamma)\, \mathcal{N}(J)\,  f(J,\gamma)\, | J,\gamma\rg\lg J,\gamma|\deq A_f \, ,
 \end{equation}
which is compatible with the energy constraint (\ref{twocond}) on the posterior distribution $J \mapsto p_n(J)$. 
Indeed, the CS quantized version of the classical Hamiltonian $H=E(J)$ is diagonal in the basis $\{|e_n\rg\, , n = 0, 1, \dotsc \}$ since it is trivially verified that in both cases the quantum Hamiltonian is exactly what we expect: 
\begin{equation}
\label{qenergy}
A_{E(J)} = \sum_{n} (E_n + \mathrm{cst}) |e_n\rg\lg e_n|\,.
\end{equation}
\end{itemize}
\subsection*{Particular quantizations}
 Actually, the quantization of any function $f_{\rm act}(J)$ of the action variable alone yields the diagonal operator:
\begin{equation} \label{(j)}
f_{\rm act}(J) \mapsto A_{f_{\rm act}} =  \sum_{n} \lg f_{\rm act} \rg_n |e_n\rg\lg e_n|\, , \quad \lg f_{\rm act} \rg_n \deq \int_{\R\, \mathrm{or}\, \R^+} d\TJ \, f_{\rm act}(J) \, p_n(J)\, .
\end{equation}
On the other hand,  the quantization of any $\tau$ periodic  function $f_{\rm ang}(\gamma)$ of the angle variable alone yields the operator:
\begin{equation} \label{qggamma1}
f_{\rm ang}(\gamma) \mapsto A_{f_{\rm ang}} =  \sum_{n,n'} [A_{f_{\rm ang}}]_{n n'} |e_n\rg\lg e_{n'}|\, , 
\end{equation}
where the matrix elements are formally given by: 
  \begin{align} 
  \label{qggamma2}
\nonumber[A_{f_{\rm ang}}]_{n n'} & = \int_{\R\, \mathrm{or}\, \R^+} d\TJ  \, \sqrt{p_n(J)\,p_{n'}(J)}\,\lim_{T \to\infty}\frac{1}{T}\int_{-\frac{T}{2}}^{\frac{T}{2}} d\gamma \, e^{-i(\alpha_n - \alpha_{n'})\gamma} \, f_{\rm ang}(\gamma)\\
 &=  \left\lbrace \begin{array}{ccc}
   0   & \mbox{if} &  \alpha_n-\alpha_{n'}\notin \frac{2\pi}{\tau} \Z\, ,  \\
    \varpi_{n n'}\, c_k({f_{\rm ang}};\tau)    &    \mbox{if}&  \alpha_n-\alpha_{n'}= \frac{2\pi}{\tau} k \in \frac{2\pi}{\tau}\Z\, ,  
\end{array}\right.
 \end{align}
 where $ c_k(f;\tau) =  \frac{1}{\tau}\int_0^{\tau}d\gamma\,f(\gamma)\, e^{-i 2\pi k\gamma/\tau}$  is the 
$k$th Fourier coefficient of $f(\gamma)$, and $ \varpi_{n n'} = \int_{\R\, \mathrm{or}\, \R^+} d\TJ  \, \sqrt{p_n(J)\,p_{n'}(J)}$  measures correlation between the two distributions $J \mapsto p_n(J)$, $J \mapsto p_{n'}(J)$. 
    Note that the diagonal values are all equal to the average of $f_{\rm ang}(\gamma)$ over one period. Also the infinite matrix can be sparse, even just diagonal, depending on the choice of the $p_n(J)$ and $\alpha_n$'s. Hence, the quantization could at end transform classical observables into a commutative algebra of operators. 
    
    An important point is that  this CS quantization procedure provides, for a given choice of the sequence $(\alpha_n)$,  a  self-adjoint angle operator $A_{\gamma}$ corresponding to the angle function $\mathcal{A}(\gamma)$ defined on the real line as the $\tau$-periodic extension of  $ \mathcal{A}(\gamma)= \gamma$ on the interval $[0, \tau)$. Then   $ c_k(\mathcal{A};\tau) = i/(2\pi k)$ for  $k\neq 0$ and  $ c_0(\mathcal{A};\tau) = \tau/2$.

 \section{The quest for ``suitable'' coherent states}
\label{expprob}  
 \subsection*{From relative error}
 Since the relative error function involves lower symbols, let us just consider those for  the two particular cases of classical functions, $ f_{\rm act} (J)$ and $f_{\rm ang} (\gamma)$. 
 \begin{align*}
\label{}
 \check{f}_{\rm act} (J)  & = \lg J,\gamma | A_{f_{\rm act}}|J,\gamma\rg= \sum_n \lg f_{\rm act}\rg_n \, \frac{p_n(J)}{\mathcal{N}(J)} \equiv \lg\lg {f_{\rm act}}\rg_n\rg_J\, ,  \\
 \check{f}_{\rm ang} (\gamma)  & =  \lg J,\gamma | A_{f_{\rm ang}}|J,\gamma\rg=\sum_{n,n'} \frac{\sqrt{p_n(J)\,p_{n'}(J)}}{\mathcal{N}(J)}\, \varpi_{n n'} c_{k(n,n')}\, e^{i\frac{2\pi}{\tau}k(n,n')\,\gamma}\,, 
\end{align*}
 where we observe, in the first case, the appearance of a double averaging using the two Bayesian facets, and, in the second case,  the presence of a deformation of the Fourier series of $f_{\rm ang}(\gamma)$ involving, on one hand, the selection rules  $k(n,n')\deq \frac{2\pi}{\tau}\,  (\alpha_n -\alpha_{n'}) \in \Z$, and weights of probabilistic origin on the other hand. The relative error (\ref{errorgen}) expresses relative deviations of the original classical observables to the above types of averaging involving the two facts of the underlying Bayesian duality. 
 
\subsection*{From localization probability distributions}
 The action-angle phase space representation of a particular coherent state  $|J_0,\gamma_0\rg$, as a function of $(J, \gamma)$,   is  the ``normalized'' overlap 
\begin{equation} \label{overdist}
\Psi_{|J_0,\gamma_0 \rg}(J,\gamma) \deq \sqrt{\mathcal{N}(J)}\,\lg J,\gamma | J_0,\gamma_0\rg = \frac{1}{\sqrt{\mathcal{N}(J_0)}}\sum_n\sqrt{p_n(J)\,p_n(J_0)}\, e^{i\alpha_n(\gamma-\gamma_0)}\, , 
\end{equation} 
 Hence,  
the map 
$X \ni (J, \gamma) \mapsto \rho^{\mathrm{phase}}_{|J_0,\gamma_0\rg}(J,\gamma) \equiv \vert \Psi_{|J_0,\gamma_0\rg}(J,\gamma)\vert^2 =\mathcal{N}(J)\,\vert \lg J,\gamma| J_0, \gamma_0\rg\vert^2$
represents a localization probability distribution, namely a generalized version of the Husimi distribution,  on the phase space provided with the pseudo-measure $\mu_B$. Indeed, the resolution of the identity gives immediately
\begin{equation} \label{}
\int_{X } \mu_B(dJ\,d\gamma)\,\rho^{\mathrm{phase}}_{|J_0,\gamma_0\rg}(J,\gamma) = 1\, .
\end{equation}
  If we choose instead a specific realization  of the Hilbert space $\mathcal{H}$, like that one generated by eigenfunctions of the quantum Hamiltonian $A_H$ in  ``$q$'' or ``configuration'' representation, $|e_n \rg \mapsto \psi_n(q)$,    the corresponding representation of the state $|J_0, \gamma_0\rg$ reads as
\begin{equation} \label{confdist}
\psi_{|J_0, \gamma_0\rg}(q) = \frac{1}{\sqrt{\mathcal{N}(J_0)}}\sum_{n} \sqrt{p_n(J_0)}\, e^{-i\alpha_n\, \gamma_0}\, \psi_n (q)\, , 
\end{equation}
with corresponding probability density of localization on the range of the $q$-variable given by $\rho^{\mathrm{circ}}_{|J_0,\gamma_0\rg}(q) \equiv \vert \psi_{|J_0,\gamma_0\rg}(q)\vert^2$

 \subsection*{From time evolution} 
Since  the CS quantized version $A_{H}$ of the classical Hamiltonian $H$ is  diagonal  in the basis $\{|e_n\rg\, ,\, n\in \Z \ (\mbox{resp.} \ \N)\, \}$, the time evolution of the CS in both representations is given respectively by
\begin{align} \label{timevJg}
\nonumber e^{-i  A_{H} t/\hbar}\Psi_{|J_0,\gamma_0\rg}(J,\gamma) &= \sqrt{\mathcal{N}(J)}\,\lg J,\gamma | e^{-iA_{H}t/\hbar}|J_0, \gamma_0\rg\\
& = \frac{1}{\sqrt{{\mathcal N} (J_0)}}\,  \sum_{n} \sqrt{p_n(J_0)p_n(J)} \,e^{i (\alpha_n(\gamma - \gamma_0)- E_n\, t/\hbar)}\,,\\
\label{timevJg}
 e^{-i  A_{H}t/\hbar}\psi_{|J_0,\gamma_0\rg}(q) &= \frac{1}{\sqrt{\mathcal{N}(J_0)}}\sum_{n} \sqrt{p_n(J_0)}\, e^{-i(\alpha_n \gamma_0 +   E_n\,t/\hbar)}\, \psi_n(q)\, . 
 \end{align}
Snapshots 
 of the time evolution of the corresponding probability densities $\vert \sqrt{\mathcal{N}(J)}\,\lg J,\gamma | e^{-i A_H t/\hbar}|J_0, \gamma_0\rg\vert^2$ and $\vert e^{-i  A_H t/\hbar}\psi_{|J_0,\gamma_0\rg}(q) \vert^2$ are necessary in order to discriminate suitable functions $p_n(J)$ and sequences $(\alpha_n)$ from the classical limit viewpoint.  Note that temporal evolution stability is  granted with the choice $\alpha_n = \alpha_{-n} = E_n$.

 \section{The quest for explicit probabilities $n \mapsto p_n(J)$: the two simple cases}
 \label{partprob}

The quantities that are left undetermined in our construction of the CS are 
the discretely indexed probability distribution $ J \mapsto p_n(J)$ and the sequence $n \mapsto \alpha_n$. 
Two simplest situations are helpful in giving some hints:  the free rotator (mass $m$ on circle of radius $l$) and the harmonic oscillator (frequency $\omega$), for which the energies are respectively $E_n \propto n^2 + $const. and $E_n \propto n + $const. 

\begin{figure}[htb!] 
\begin{center}
\includegraphics[width=3in]{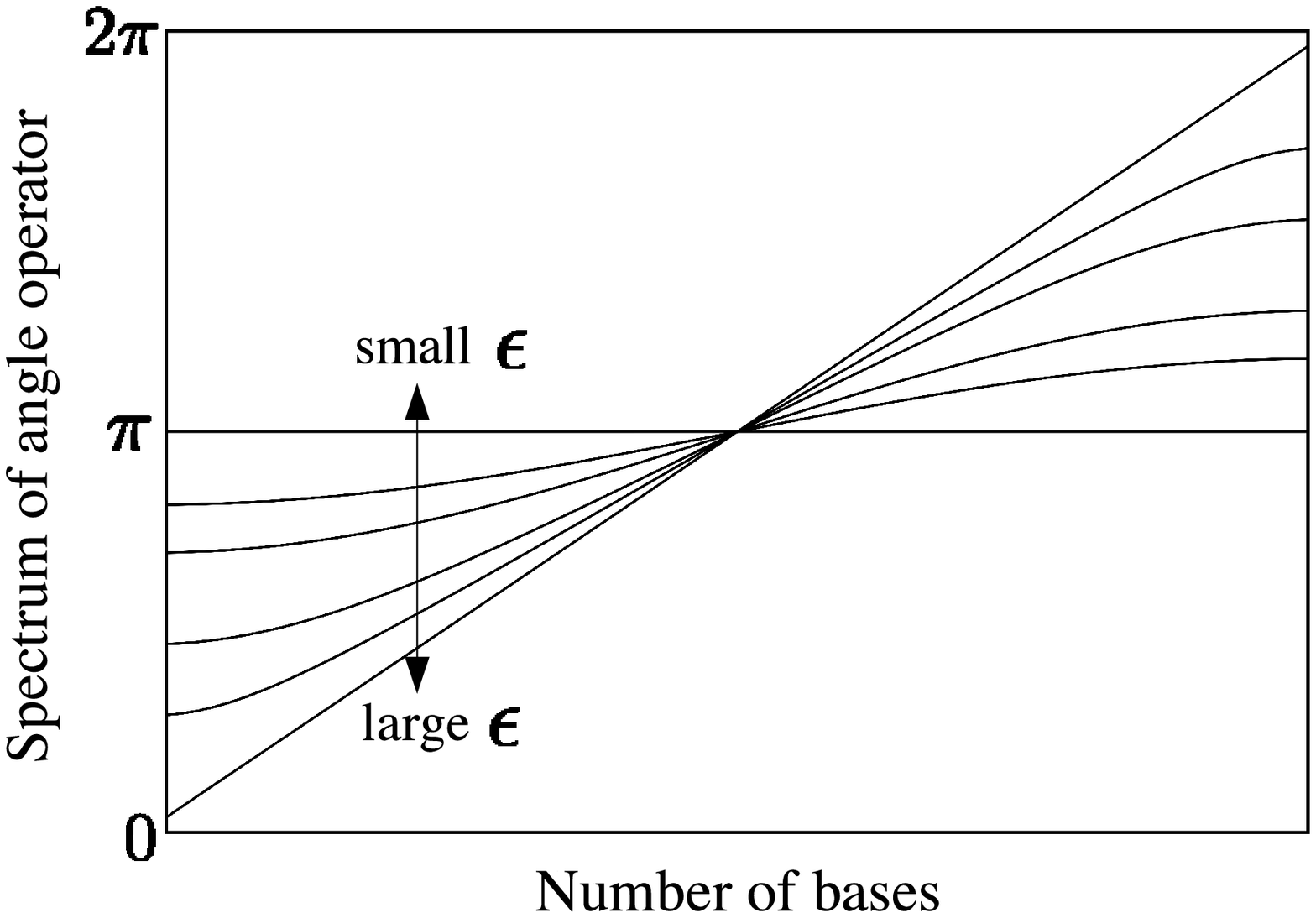}
\includegraphics[width=3in]{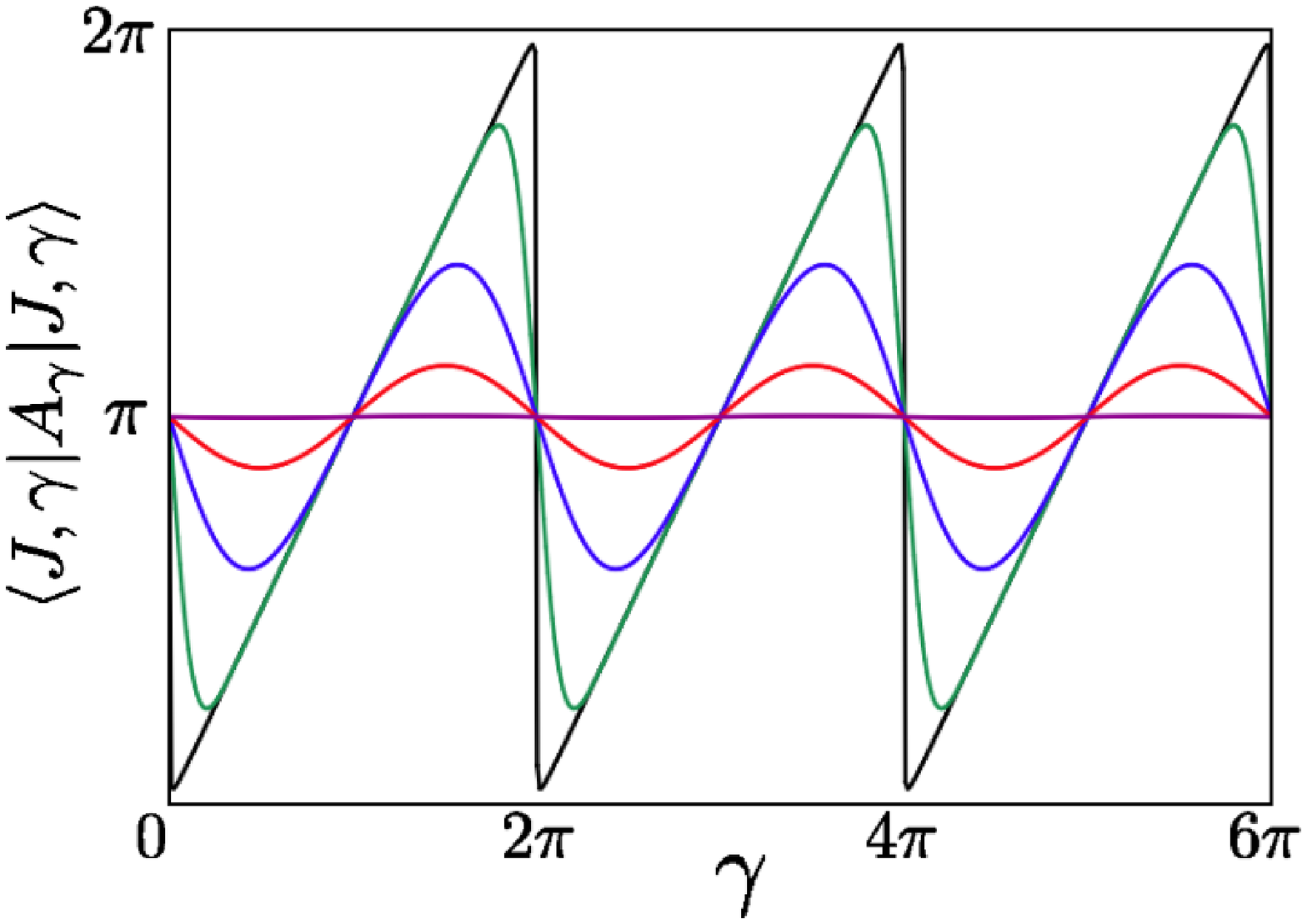}
\caption{Spectrum (left) and lower symbol (right) of the angle operator obtained by CS quantization of the angle of rotation on the circle, for different values of parameter $\epsilon= 1/(2\sigma^2)$, when $p_n(J)=   \left(\frac{\epsilon}{\pi}\right)^{1/2}\,e^{-\epsilon(\TJ- n)^2 }$ and $\alpha_n = n$.
For the spectrum,  $\epsilon = 10^{-7}, 0.3, 1, 3, 5, 50$.
For the lower symbol: $\epsilon = 10^{-7}, 0.1, 1, 3, 10$   (the lower symbol has little dependence on $J$).
One notices the tendency to  the sawtoothed behavior  of the classical angle function as $\epsilon \to \infty$, i.e. as $\sigma \to 0$.}
\label{specg}
\end{center}
\end{figure}

In the first case,  a familiar solution \cite{cscircle} is family of Gaussians 
centered at each integer, with dimensionless width parameter $\sigma$ or $\epsilon\equiv 1/(2\sigma^2)$:
\begin{equation} 
\label{gaussrot}
p_n(J)= \left(\frac{1}{2\pi\sigma^2}\right)^{1/2}\,e^{-\frac{1}{2\sigma^2h^2}(J-h n)^2 }\equiv   \left(\frac{\epsilon}{\pi}\right)^{1/2}\,e^{-\epsilon(\TJ- n)^2 } \, , \quad n\in \Z \, ,
\end{equation}
with $\tilde{J} = J/h$. 
This gives the eigenvalues $J_n = h n$ of $A_J$ and 
\begin{equation}
\label{freerotene}
E_n = \frac{h^2 n^2}{8 \pi^2 m l^2} +  \frac{\sigma^2 h^2}{8\pi^2 m l^2}=\frac{h^2 n^2}{8 \pi^2 m l^2} +  \frac{h^2}{16\epsilon \pi^2 m l^2}\, , 
\end{equation}
the constant shift being the average value of the classical energy with respect to the distribution $p_0(J)$. 
By introducing the Compton length $\lambda_c = \hbar/mc$ of the particle Eq. (\ref{freerotene}) reads $E_n = (\lambda_c/l)^2mc^2(n^2 + 1/(2\epsilon))/2$.  
As an illustration of the suitability of this Gaussian choice, we show in Figure \ref{specg} the respective behaviors of the spectrum and  lower symbol of the angle operator $A_{\gamma}$ for different values of the parameter $\epsilon$. 

  In the second case, another familiar solution is the discretely indexed gamma distribution:
\begin{equation} \label{}
p_n(J)= e^{-\tilde J}\, \frac{\tilde J^n}{n!}\, , \quad n\in \N \, . 
\end{equation}
This gives $J_n = h (n+1)$ and $E_n = \dfrac{h \,\omega }{2 \pi}(n  +  1)= \hbar\omega(n+1)$, the constant shift being the average value of the classical energy with respect to the distribution $p_0(J)$.

 \section{The quest for explicit probabilities: the general case}
\label{genprob}
Given the classical relation $E=E(J)$ between action variable and energy, and  the observational or computed sequence $(E_n)$,  the central question is to find the sequence of probability distributions  $J\mapsto p_n(J)$ (at least with a satisfying approximation),  which obey the two fundamental conditions (\ref{twocond}). 
 For the rotation case, the departure point could be a normal-like law, possibly modified along a perturbation scheme with expansion parameter the strength of the potential energy $U$. Similarly, in the libration case, the departure could be a gamma-like law, possibly modified along the same lines. 
A nice and manageable model is  the  simple pendulum, whose the quantum version  is well known from the solution of  the Mathieu equation \cite{abra}.

 \subsection*{Approximations for the simple pendulum: the rotation case}
  
  Starting from the computed Mathieu eigenvalues $E_n$,  an empirical approach
consists in starting from the sequence  of computed action variables $J^{\mathrm{cl}}_n \deq J(E_n)$ from Eq. (\ref{action1}) and to impose,
 in the rotation case, the sequence of  normal laws centered at  these $J^{\mathrm{cl}}_n$.
  \begin{equation} \label{}
p_n(J)= \left(\frac{1}{2\pi\sigma_n^2}\right)^{1/2}\,e^{-\frac{1}{2\sigma_n^2h^2}(J- J^{\mathrm{cl}}_n)^2 }\, , \quad n\in \Z \, , 
\end{equation}
by ``adjusting''  $\sigma_n$ in order to suitably approximate  the $E_n$'s with the computed quantities 
 \begin{equation} \label{}
E_n^{\mathrm{ app}} + \mathrm{cst}\deq\int_{-\infty}^{+\infty} E(J)\, p_n(J)\,d\TJ\, . 
\end{equation}
  Note that with this choice, the eigenvalues $J_n$ of $A_J$ are precisely the $J^{\mathrm{cl}}_n$'s. 
  \subsection*{Approximations for the simple pendulum: the libration case}
  
Handling the libration case is more delicate. Another empirical approach
consists  in introducing the gamma-like distribution:
 \begin{equation*}
J \mapsto p_n(J)= \frac{1}{ \mathcal{E}_y(J)}\, \frac{\TJ^{n}}{y_n!}\, , \quad n\in \N \, , \quad y_n!\deq y_1 y_2 \cdots y_n , \quad  y_0!\deq 1\, , \quad \mathcal{E}_y(J)\deq\sum_{n=0}^{\infty}   \frac{{\TJ}^n}{y_n!}\, ,
\end{equation*}
with an auxiliary sequence $ \{0= y_0 < y_1< \cdots < y_n < \cdots\}$ such that the corresponding 
moment problem has a solution with a positive measure $w_y(J)\, d\TJ$,
\begin{equation}
\label{momham}
\int_{0}^{+\infty}  d\TJ\, w_y(J)\, p_n(J) = 1\, , 
\end{equation}
\underline{and} that   the quantization conditions involving the computed Mathieu eigenvalues, 
 \begin{equation*}
E_n + \mathrm{cst}=  \int_{0}^{+\infty}d\TJ\, w_y(J)\, E(J)\, p_n(J)\, , \quad n=0, 1,2,\cdots\, , 
\end{equation*} 
are fulfilled (at least approximately). 
We note that, by construction, we have $\sum_{n=0}^{\infty} p_n(J) = 1$,  that the action variable $J$ is the mean value of $n\mapsto y_n$ with respect to the Poisson-like distribution $n\mapsto p_n(J)$: $\lg y_n\rg \deq \sum_{n=0}^{\infty} y_n\, p_n(J) = J$, and that the quantum action $A_J$ has eigenvalues  $\lg J\rg_J = y_{n+1}$. 

\section{Free rotor CS: two interesting choices} 
 \label{timevfr}
 With the Gaussian choice (\ref{gaussrot}) and for a general choice of a sequence of frequencies $\alpha_n$, the coherent states for the free rotator read 
\begin{equation}
\label{ccs15}
 | J, \gamma \rangle =  \frac{1}{\sqrt{\mathcal{N} (J)}}\,  \left(\frac{\epsilon}{\pi}\right)^{1/4} \sum_{n \in \Z} e^{-\frac{\epsilon}{2}(\TJ-n)^2 } \,e^{- i \alpha_n \gamma} | e_n \rangle\, ,
\end{equation}
 The normalization function $\mathcal {N} (J) $ is given in two forms:
\begin{equation} \label{}
\mathcal{ N} (J) = \sqrt{\frac{\epsilon}{\pi}}\sum_{n \in \Z} e^{-\epsilon (\TJ-n)^2} \underset{\mbox{Poisson}}{=}  \sum_{n \in \Z} e^{2\pi i n\TJ}\, e^{-\frac{\pi^2}{\epsilon} n^2}\, ,
\end{equation}
and satisfies $\lim_{\epsilon \to 0}\mathcal{N}(J) = 1$. 
 In the following we consider two choices of $\alpha$ and investigate what kind of properties of CS is satisfied for each case. 

\subsection*{Case $\alpha_n = 2\pi n/\tau$} 

This choice  renders exact the quantization of the classical canonical  commutation rule $\left\{ J, e^{i2\pi \gamma/\tau} \right\} = i e^{i2\pi \gamma/\tau}$. Indeed, we have $\lbrack A_J, A_{e^{i2\pi \gamma/\tau}} \rbrack = h\,A_{e^{i2\pi \gamma/\tau}}$. 
Concerning the phase space distribution (\ref{evphasdist}) for time evolution, we obtain the  
 following upper bound:
\begin{equation} \label{upboundn}
\rho_{J_0,\gamma_0}(J,\gamma;t)\leq \frac{\epsilon}{\sqrt{\epsilon^2 + \tilde{t}^2}}\,  \frac{e^{-\epsilon \frac{(\TJ-\TJ_0)^2}{2}}}{{\mathcal N} (J_0)}\sum_{n \in \Z} e^{-\frac{\epsilon}{2(\epsilon^2 + \tilde{t}^2)}\,\left(2\pi n -\tilde{\gamma} +\mu  \tilde{t} \right)^2}  \, ,
\end{equation}
where $\epsilon = 1/(2\sigma^2)$,  $\mu = (\TJ+\TJ_0)/2$, $\tilde{t}= \hbar t/(2ml^2)$, and
$\tilde{\gamma} = 2\pi(\gamma -\gamma_0)/\tau$.
 From this is derived the estimate on  the semi-classical behavior  at large $\TJ_0 = M \in \Z$:
$\rho_{J_0,\gamma_0}(J,\gamma;t) \leq \frac{1}{\sqrt{1 + 4\sigma^4\tilde{t}^2}}\, \delta_{\TJ \TJ_0}$
for $ \tilde{\gamma}/2\pi - M\tilde{t} \in \Z\, , $
and vanishes if $\TJ_0 \notin \Z$ or $ \tilde{\gamma}/2\pi - M\tilde{t} \notin \Z$.

 \subsection*{Case $\alpha_n = 2\pi n^2/\tau$}
 This choice is  appropriate  to  temporal evolution stability  
  \begin{equation}
\label{evopfreerot}
e^{-iA_H t/\hbar} |J,\gamma\rg = e^{-i\tilde{t}/2} \left| J, \gamma - \frac{\tau }{2\pi}\tilde{t}\right\rg\, . 
\end{equation}
This time, the phase space distribution (\ref{evphasdist}) is bounded as follows. 

$\lambda = \frac{2\pi}{\tau}(\gamma -\gamma_0 -\hbar t/(4\pi m l^2))$
\begin{equation} \label{upboundn2}
\rho_{J_0,\gamma_0}(J,\gamma;t)  \leq  \frac{\epsilon}{\sqrt{\epsilon^2 + \tilde\gamma^2(t)}}\,  \frac{e^{-\epsilon\frac{(\TJ-\TJ_0)^2}{2}}}{{\mathcal N} (J_0)}\sum_{n \in \Z} e^{-\frac{\epsilon}{2(\epsilon^2 + \tilde\gamma^2(t))}\,\left(2\pi n -2 \mu \tilde\gamma(t) \right)^2}\, , 
\end{equation}
where $\tilde\gamma(t) =  \frac{2\pi}{\tau}(\gamma -\gamma_0 -\hbar t/(4\pi m l^2)) = \tilde\gamma - \tilde t$. 
Hence the estimate on semi-classical behavior of $\rho_{J_0,\gamma_0}(J,\gamma;t) $ at large $\TJ_0 = M \in \Z$:
$\rho_{J_0,\gamma_0}(J,\gamma;t) \leq  \frac{1}{\sqrt{1 + 4\sigma^4\tilde\gamma^2(t)}}\, \delta_{\TJ \TJ_0}$
for $ \tilde\gamma(t) \in 2\pi\Z/M\, , $
and vanishes if $\TJ_0 \notin \Z$ or $\tilde\gamma(t) \notin 2\pi\Z/M$.

 \section{Concluding points}
 \label{concl}

Periodic and more generally  Integrable systems 
 provide a variety of such families of action-angle coherent states.  
  A future issue is about the best choice of 
probability distributions $n_i \mapsto p_{n_i} (J_i)$. A more fundamental
concerns the physical (in terms of physical measurement) equivalence between different families of coherent states from  a quantization point of view. Finally, the extension of the probabilistic approach presented in this work to 
unconfined systems 
and subsequent continuous spectra is possible \cite{bagagily}.

\end{document}